# Computer Simulation of Neural Networks Using Spreadsheets: The Dawn of the Age of Camelot

Serhiy O. Semerikov[1][0000-0003-0789-0272], Illia O. Teplytskyi[1],
Yuliia V. Yechkalo[2][0000-0002-0164-8365] and Arnold E. Kiv[3]

[1] Kryvyi Rih State Pedagogical University, 54, Gagarina Ave., Kryvyi Rih, 50086, Ukraine
semerikov@gmail.com
[2] State Institution of Higher Education "Kryvyi Rih National University",
11, Vitali Matusevich St., Kryvyi Rih, 50027, Ukraine
uliaechk@gmail.com
[3] Ben-Gurion University of the Negev, Beer Sheba, Israel
kiv@bgu.ac.il

**Abstract.** The article substantiates the necessity to develop training methods of computer simulation of neural networks in the spreadsheet environment. The systematic review of their application to simulating artificial neural networks is performed. The authors distinguish basic approaches to solving the problem of network computer simulation training in the spreadsheet environment, joint application of spreadsheets and tools of neural network simulation, application of third-party add-ins to spreadsheets, development of macros using the embedded languages of spreadsheets; use of standard spreadsheet add-ins for non-linear optimization, creation of neural networks in the spreadsheet environment without add-ins and macros. After analyzing a collection of writings of 1890-1950, the research determines the role of the scientific journal "Bulletin of Mathematical Biophysics", its founder Nicolas Rashevsky and the scientific community around the journal in creating and developing models and methods of computational neuroscience. There are identified psychophysical basics of creating neural networks, mathematical foundations of neural computing and methods of neuroengineering (image recognition, in particular). The role of Walter Pitts in combining the descriptive and quantitative theories of training is discussed. It is shown that to acquire neural simulation competences in the spreadsheet environment, one should master the models based on the historical and genetic approach. It is indicated that there are three groups of models, which are promising in terms of developing corresponding methods – the continuous two-factor model of Rashevsky, the discrete model of McCulloch and Pitts, and the discrete-continuous models of Householder and Landahl.

# 1 Introduction

For the past 25 years, the authors have been developing the concept of systematic computer simulation training at schools and teachers' training universities [53]. The concept ideas have been generalized and presented in the textbook [60]. Spreadsheets are chosen to be the leading environment for computer simulation training [52; 71], their application discussed in articles [55; 62; 66; 68]. Using spreadsheet processors (autonomous [18], integrated [61] and cloud-oriented [72]) as examples, the authors demonstrate components of teaching technology of computer simulation [70] of determined and stochastic [56; 58; 67] objects and processes of various nature [59; 69].

The systematic training of simulation provides for changing [52] and integrating [57] simulation environments ranging from general (spreadsheets) to specialized subject-based ones. While teaching computer simulation of intellectual systems [50] specialized languages and programming environments [28] are traditionally used. They can be easily mastered by first-year students [1; 27]. One of the most wide-spread languages, Scheme, is offered to be applied to teaching computer simulation of classical mechanics at universities [54]. Extensive application of artificial intelligence in everyday life calls for students' early acquaintance with its models and methods including neural network-based [29] while teaching informatics at secondary schools. It conditions the need for developing training methods of computer simulation of neural networks in the general-purpose simulation environment, i.e. spreadsheets.

# 2 Literature Review and Problem Statement

The first description of spreadsheet application to teaching neural network simulation of visual phenomena dates back to 1985 and belongs to Thomas T. Hewett, Professor of the Department of Psychology of Drexel University [11]. In [10] there are described simple models of microelectrode recording of two neuron types of neural activity – receptors and transmitters localized in two brain-hemispheres. Thomas T. Hewett offered psychology students to independently choose coefficients of intensifying or reducing input impulses to achieve the desired output: "... the simulations can be designed in such a way that the student is able to "experiment" with a simulation-experiment both in the sense of discovering the characteristics of an unknown model and in the sense of modifying various components of a known model to see how the simulation is affected" [10, p. 343]. This approach implies simultaneous studying a neural network and understanding its functioning as psychology students conclude the laws of the neural impulse spread by applying the trial-and-error method.

In his article [4], James J. Buergermeister, Professor of Hospitality and Tourism Management of University Wisconsin-Stout, associates electronic spreadsheet application with basic principles of training technology and methods of data processing (Fig. 1). The author does not work out the methods of applying electronic spreadsheets to neural network simulation in detail, yet, the presented scheme reveals such basic steps as data obtainment, semantic coding, matching with an etalon, etc.

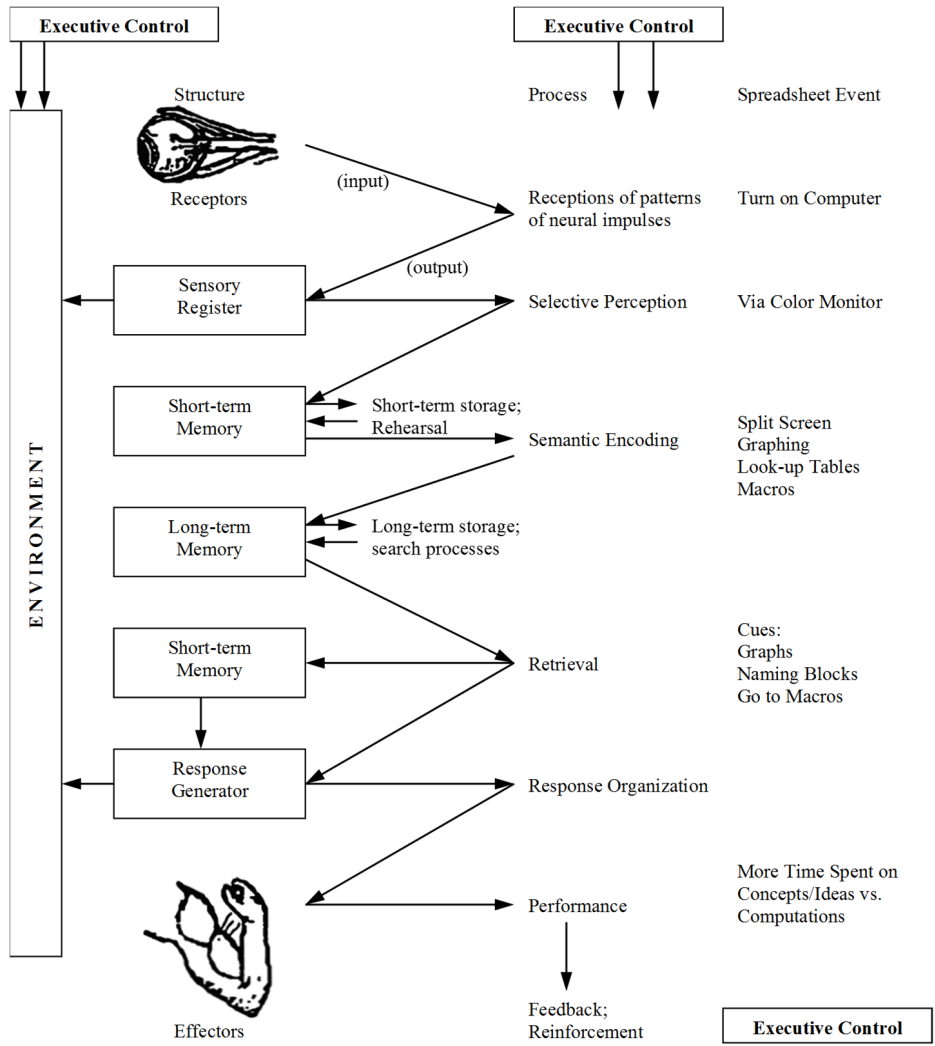

**Fig. 1.** The information-processing model using spreadsheet events (according to [4])

Since 1988, Murray A. Ruggiero, one of the pioneers of autotrading, has been developing Braincel, an application for Microsoft Excel 2.1C, which is a set of twenty macros to solve tasks of image recognition by artificial neural network tools [21]. At the beginning of 1991, Murray A. Ruggiero received a patent "Embedding neural networks into spreadsheet applications" [49], which describes an artificial neural network with a plurality of processing elements called neurons arranged in layers. They further include interconnections between the units of successive layers. A network has an input layer, an output layer, and one or more "hidden" layers in between, necessary to allow solutions of non-linear problems. Each unit (in some ways analogous to a biological neuron: dendrites – input layer, axon – output layer, synapses – weights [47], soma – summation function) is capable of generating an output signal which is determined by the weighted sum of input signals it receives and an activation function specific to that unit. A unit is provided with inputs, either from outside the network or from other units, and uses these to compute a linear or non-linear output. The unit's output goes either to other units in subsequent layers or to outside the network. The input signals to each unit are weighted by factors derived in a learning process.

When the weight and activation function factors have been set to correct levels, a complex stimulus pattern at the input layer successively propagates between the hidden layers, to result in a simpler output pattern. The network is "taught" by feeding it a succession of input patterns and corresponding expected output patterns. The network "learns" by measuring the difference at each output unit between the expected output pattern and the pattern that it just produced. Having done this, the internal weights and activation functions are modified by a learning algorithm to provide an output pattern which most closely approximates the expected output pattern, while minimizing the error over the spectrum of input patterns. Neural network learning is an iterative process involving multiple lessons. Neural networks have the ability to process information in the presence of noisy or incomplete data and yet still generalize to the correct solution.

In his patent, Murray A. Ruggiero details a network structure (multi-level), an activation function (sigmoidal), a coding method (polar), etc. He presents a mathematical apparatus for network training and determines a method of data exchange between a spreadsheet processor nucleus and an add-in to it. The patent author suggests storing input data in columns, maximum and minimum values for each column of input data, the number of learning patterns. Data can be normalized or reduced to the polar range [0; 1] both in spreadsheets and add-ins.

In his article of 1989, Paul J. Werbos, the pioneer of the backpropagation method for artificial neural network training [65] demonstrates how to make the corresponding mathematical apparatus simpler to use it directly in the spreadsheet processor. The cycling character of training is supported by a macro that exchanges data between lines to avoid restrictions on the number of iterations because of the limited number of lines on a sheet of a separate spreadsheet. Some other authors suggest applying a similar approach of macros application [8; 74].

The authors of [22] in Chapter 2 "Neural Nets in Excel" give an example of applying the non-linear optimization tool, Microsoft Excel Solver, to forecasting stock prices using the "grey-box" concept, in which the model is evident, yet, the details of its realization are hidden.

In their article of 1998 [9], Tarek Hegazy and Amr Ayed from the Department of Civil Engineering at University of Waterloo distinguish the corresponding seven steps (Fig. 2). Unlike [48], the authors suggest using bipolar data normalization (over the range of [–1; 1]) and a hyperbolic tangent as an activation function. Three add-ins for Microsoft Excel are used to determine weighting factors – the standard Solver and third-party add-ins (NeuroShell2 and GeneHunter by Ward Systems Group). Experiment results reveal that the best result is provided by the optimizing general-purpose tool (Solver) and not by specialized ones. In spite of the fact that "Journal of Construction Engineering and Management" does not refer to educational editions, the article

[9] and the paper [3] by their structure and focus on details can be considered the first description of methodic of using spreadsheets for neural network simulation.

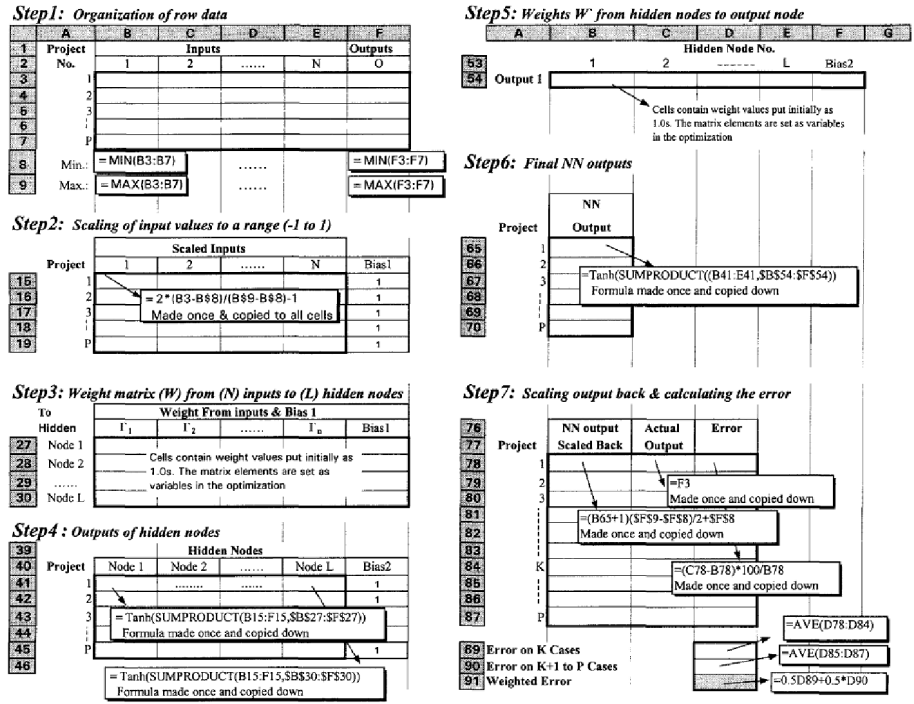

**Fig. 2.** Spreadsheet simulation of three-layer neural network with one output node (according to [9])

In their article of 2012 [47], Thomas F. Rienzo and Kuriakose K. Athappilly from Haworth College of Business at Western Michigan University consider model illustrating the process of machine learning as networks examine training data would provide another. Authors incorporate the stepwise learning processes of artificial neural network in a spreadsheet containing (1) a list or table of training data for binary input combinations, (2) rules for target outputs, (3) initial weight factors, (4) threshold values, (5) differences between target outputs and neural network transformation values, (6) learning rate factors, and (7) weight adjustment calculations. Unlike the previous ones, this model is invariant to the spreadsheet and does not call for applying any third-party add-ins.

The conducted review makes it possible to find the following solutions of the problem of computer simulation teaching to neural networks in the spreadsheet environment:

— joint application of spreadsheets and neural network tools [29], in which data is exported to the unit calculating weighting factors imported to spreadsheets and used in calculations;

- application of third-party add-ins for spreadsheets ([9; 21; 37; 49]), according to which structured spreadsheet data is processed in the add-in, calculation results are arranged in spreadsheet cells;
- macros development ([3; 8; 65; 74]) enables direct software control over neural network training and creation of a user's specialized interface;
- application of standard add-ins for optimization ([9; 22; 37]) calls for transparent network realization and determination of an optimization criterion (minimization of a squared deviation total of the calculated and etalon outputs of the network);
- creation of neural networks in the spreadsheet environment without add-ins and macros [47] requires transparent realization of a neural network with evident determination of each step of adjustment of its weighting factors.

The advantage of the first solution is its flexibility as one can choose any relevant combinations of the simulation environments, yet, their integration level is usually insufficient. The closed character of the second solution and its binding to a certain software platform make it relevant to be applied to solving various practical tasks and irrelevant for neural network simulation training as a network becomes a black box for a user. The fourth solution is partially platform-dependent as a neural network becomes a grey box for a user. The final solution is totally mobile and offers an opportunity to regard the model as a white box, thus making it the most relevant for initial mastering of neural network simulation methods.

## 3  The Aim and Objectives of the Study

The research is aimed at considering mathematical models of neural networks realized in spreadsheet environment.

To accomplish the set goal, the following tasks are to be solved:

1. to distinguish learning tools of computer simulation of neural networks in the spreadsheet environment;
2. to study mathematical models of neural networks at the beginning of the Age of Camelot [7].

## 4  Mathematical models of neural networks at the beginning of the Age of Camelot (1933-1947)

Russell C. Eberhart and Roy W. Dobbins [7] suggest dividing the history of artificial network development into four stages. The first stage, the Age of Camelot, starts with "The Principles of Psychology" (1890) by the American psychologist, William James, who formulates the elementary law of association: "When two elementary brain processes have been active together or in immediate succession, one of them, on re-occurring, tends to propagate its excitement into the other" [20, p. 566]. The elementary law of association (the elementary principle) is closely related to the concepts of associative memory and correlational learning. In the authors' opinion [7], William James seemed to foretell the notion of a neuron's activity being a function of the sum of its inputs, with past correlation history contributing to the weight of interconnections: "The amount of activity at any given point in the brain-cortex is the sum of the tendencies of all other points to discharge into it, such tendencies being proportionate (1) to the number of times the excitement of each other point may have accompanied that of the point in question; (2) to the intensity of such excitements; and (3) to the absence of any rival point functionally disconnected with the first point, into which the discharges might be diverted" [20, p. 567].

William James illustrates his elementary principle by total recall example: "Suppose, for example, we begin by thinking of a certain dinner-party. The only thing which all the components of the dinner-party could combine to recall would be the first concrete occurrence which ensued upon it. All the details of this occurrence could in turn only combine to awaken the next following occurrence, and so on. If *a, b, c, d, e*, for instance, be the elementary nerve-tracts excited by the last act of the dinner-party, call this act **A**, and *l, m, n, o, p* be those of walking home through the frosty night, which we may call **B**, then the thought of **A** must awaken that of **B**, because *a, b, c, d, e*, will each and all discharge into *l* through the paths by which their original discharge took place. Similarly they will discharge into *m, n, o,* and *p*; and these latter tracts will also each reinforce the other's action because, in the experience **B**, they have already vibrated in unison. The lines in ... [Fig. 3] symbolize the summation of discharges into each of the components of **B**, and the consequent strength of the combination of influences by which **B** in its totality is awakened." [20, p. 569].

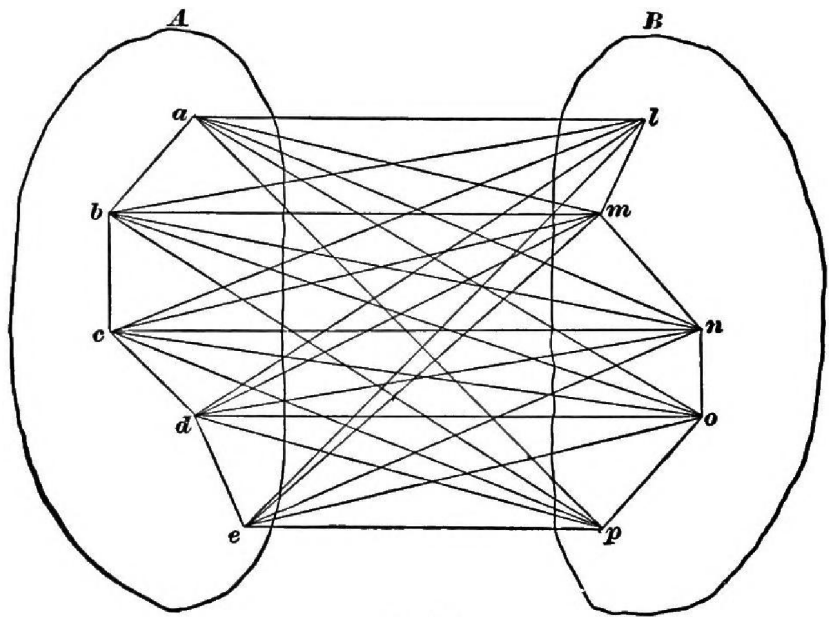

**Fig. 3.** Thinking acts according to William James [20, p. 570]

Fig. 3 reveals the neural network depicting connection between two conditions. Applying the elementary principle to analyzing forgetting and recall, William James creates an associative neural network (Fig. 4): "The whole process can be rudely symbolized in a diagram. Call the forgotten thing **Z**, the first facts with which we felt it was related *a*, *b*, and *c*, and the details finally operative in calling it up, *l*, *m*, and *n*. Each circle will then stand for the brain-process principally concerned in the thought of the fact lettered within it. The activity in **Z** will at first be a mere tension; but as the activities in *a*, *b*, and *c* little by little irradiate into *l*, *m*, and *n*, and as all these processes are somehow connected with **Z**, their combined irradiations upon **Z**, represented by the centripetal arrows, succeed in rousing **Z** also to full activity. ... Turn now to the case of finding the unknown means to a distinctly conceived end. The end here stands in the place of *a*, *b*, *c*, in the diagram. It is the starting-point of the irradiations of suggestion; and here, as in that case, what the voluntary attention does is only to dismiss some of the suggestions as irrelevant, and hold fast to others which are felt to be more pertinent – let these be symbolized by *l*, *m*, *n*. These latter at last accumulate sufficiently to discharge all together into **Z**, the excitement of which process is, in the mental sphere, equivalent to the solution of our problem. The only difference between this case and the last, is that in this one there need be no original sub-excitement in **Z**, cooperating from the very first. When we seek a forgotten name, we must suppose the name's centre to be in a state of active tension from the very outset, because of that peculiar feeling of recognition which we get at the moment of recall." [20, pp. 586-588].

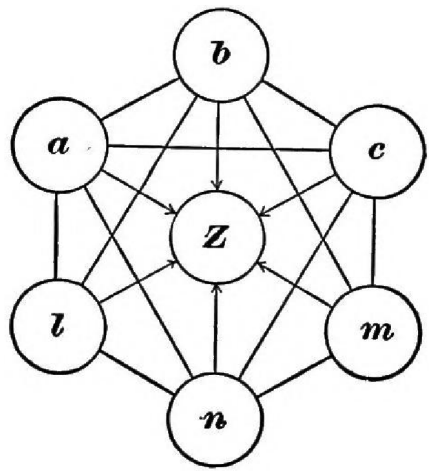

**Fig. 4.** The recall process according to William James [20, p. 586]

In "Psychology" (1892), an abridged re-edition of "The Principles of Psychology", William James formulates basic principles of the image recognition theory: "We know, in short, a lot about it, whilst as yet we have no acquaintance with it. Our perception that one of the objects which turn up is, at last, our quæsitum, is due to our recognition that its relations are identical with those we had in mind, and this may be a rather slow act of judgment. Every one knows that an object may be for some time present to his mind before its relations to other matters are perceived. Just so the relations may be there before the object is." [19, p. 275].

"The Bulletin of Mathematical Biophysics" has been an advanced platform for approbating network models and methods since the moment of its foundation by Nicolas Rashevsky in 1939 [6]. It should be no surprise as Nicolas Rashevsky invented one of the first models of the neuron [44] and started the idea of artificial neural networks. The basic idea was to use a pair of linear differential equations and a nonlinear threshold operator (1):

$$\text{Input} := I(t)$$
$$\begin{cases} \dfrac{de}{dt} = AI(t) - ae \\ \dfrac{dj}{dt} = BI(t) - bj \end{cases} \quad (1)$$
$$\text{Output} := H(e - j - \theta)$$

where $\theta$ is the threshold, $e$ and $j$ could represent excitation and inhibition or the amount or concentration of two substances within a neuron, $H(x)$ is the Heaviside operator (takes positive values to 1, and non-positive values to 0). This gives an easy way to model the all-or-none firing of a neuron – Nicolas Rashevsky showed that this simple model was able to model many of the known experimental results for the behavior of single neurons. He also made the point that networks of these model neurons could be connected to give quite complicated behavior and even serve as a model for a brain [6].

In his article of 1941, Gale J. Young ("the best theoretical engineer" of Manhattan Project [64]) [73], shows that the Nicolas Rashevsky two-factor model of nerve excitation can account for sustained inhibition or enhancement by a sequence of stimulus pulses, and for the decrease in the reinforcement period with each successive pulse of the train.

Developing Nicolas Rashevsky's ideas, his student Alston Scott Householder, who gave his name to the known linear transformation describing a reflection about a plane or hyperplane containing the origin, and a class of root-finding algorithms used for functions of one real variable with continuous derivatives up to some order, in his article of 1940 [13], suggests a parameter measuring the "strength" of the inhibitory neurons acting among the terminal synapses. In [16], he describes the activity parameter as a characteristic of the fiber which is assumed to be different from zero, but it may be either positive (when the fiber is excitatory in character) or negative (when the fiber is inhibitory in character). In Householder's articles of 1941–1942:

— Preliminary consideration is given to the steady-state activity of some simple neural structures. It is assumed as a first approximation that while acted upon by a constant stimulus, each fiber reaches a steady-state activity whose intensity is a linear function of the applied stimulus. It is shown by way of example that for a simple two-fiber circuit of inhibitory neurons knowledge of the stimuli applied to the separate fibers does not necessarily suffice to determine uniquely the activity that will result. On the other hand, there are deduced certain restrictions on the possible types of activity that may be consistent with a given pattern of applied stimulation [16].

- It is found that for a simple circuit of neurons, if this contains an odd number of inhibitory fibers, or none at all, or if the product of the activity parameters is less than unity, then the stimulus pattern always determines uniquely the steady-state activity. For circuits not of one of these types, it is possible to classify exclusively and exhaustively all possible activity patterns into three types, here called "odd", "even", and "mixed". For any pattern of odd type and any pattern of even type there always exists a stimulus pattern consistent with both, but in no other way can such an association of activity patterns be made [17].
- It is shown here that when the product of the activity parameters of the neural circuit is not exceeded by unity (algebraically) a steady state is not possible in which all fibers of the circuit are active, whereas when this product is exceeded by unity, any stimulus pattern which is consistent with such a state of complete activity is inconsistent with any state of partial activity of the circuit [14].
- Conditions under which either of two distinct activity patterns may arise from the same stimulus pattern are deduced for the case of a network which consists of $N$ simple circuits all jointed at a common synapse. If the product of the activity parameters of all the fibers in any circuit is called the activity parameter of the circuit, or, more briefly, the circuit parameter, then the condition for the existence of such mutually consistent activity patterns is that there be a sum of circuit parameters which is not less than unity [15].

Thus, at the beginning of 1942, the theory of biological neural networks based on Nicolas Rashevsky's continuous two-factor model was created and intensively developed. As remembered by J. A. Anderson and E. Rosenfeld, at the boundary of two decades, Walter Pitts was introduced to Nicolas Rashevsky by Rudolf Carnap, and accepted in to his mathematical biology group [5]. In his early publication, Walter Pitts suggests "a new point of view in the theory of neuron networks is here adumbrated in its relation to the simple circuit: it is shown how these methods enable us to extend considerably and unify previous results for this case in a much simpler way" [33, p. 121]. With due consideration of Householder's articles, Walter Pitts determines the total conduction time of a fiber as the sum of its conduction time and the synaptic delay at the postliminary synapse. Walter Pitts was the first to use spreadsheet abstraction and discrete description of neural network functioning by determining a corresponding algorithm: "The excitation-pattern of [neural circuit] $C$ may be described in a *matrix E*, of $n$ rows and an infinite number of columns, each of whose elements $e_{rs}$ represents the excitation at the synapse $s_r$ during the interval $(s, s+1)$. The successive entries in the excitation matrix $E$ may be computed recursively from those in its first column – these are the quantities $\lambda_r$ – by the following rule, whose validity is evident: Given the elements of the $p$-th column, compute those of the $p$+l-st thus: if the element $e_{ip}$ is negative or zero, place $\sigma_{i+1}$ in the $i$+l-st row and $p$+1-st column, or in the first row of the $p$+l-st column if $i=n$. Otherwise put $\sigma_{i+1}+a_i e_{ip}$, in this place. We shall say that $C$ is in a *steady-state* during a series of $n$ intervals $(s, s+1), ..., (s+n–1, s+n)$ if, for every $p$ between $s$ and $s+n$, the $p$-th and $p+n$-th columns of $E$ are identical. If $s$ is the smallest integer for which this is the case, we shall say that the steady state begins at the interval $(s, s+1)$" [33, pp. 121–122]. Rather than analyzing the steady-state activity of networks, Walter Pitts was more concerned with initial nonequilibrium cases, and how a steady state could be achieved [2, p. 18].

The suggested algorithm describes a parallel neural network: "It will be seen that the construction of the matrix $E$ implies that its infinite diagonals – where we take a diagonal to start again at the top of the succedent column whenever it reaches the last row of $E$ – are wholly independent of one another, so that if we know the starting point of a diagonal of $E_s$, we can calculate the entries along it uncognizant of any other values in the matrix. Physically, this of course means that the activity in $C$ can be regarded as composed of wholly independent impulses, commencing originally at a synapse $s_j$ with a value $\lambda_j$, and journeying around $C$ in irrelation to the impulses beginning at other synapses. We shall find it convenient to adopt this standpoint, and consider only the case of a single impulse, so that the complete solution must be derived by combining the results of our subsequent procedures for the separate diagonals, and a steady-state for the whole circuit is attained only when one has accrued for each separate diagonal [33, p. 122].

The results provided by Walter Pitts in his articles on the linear theory of neuron networks (the static problem [35] and the dynamic problem [34]), enabled him to draw two essential conclusions: (1) it is possible to find a set of independent networks each of which consists of *n* simple circuits with one common synapse (*rosettes*), such that network arises by running chains from the centers of the rosettes to various designated points outside: but none back, so that the state of the whole network is determined by the states of the separate rosettes independently – Pitts calls networks of this kind *canonical networks* [34, p. 29]; (2) given any finite network, it is possible to find a set of independent rosettes such that the excitation function of network for every region is a linear combination of those of the rosettes – i. e., we can reduce any network to a canonical network having the same excitation function [34, p. 31]. Thus, in his article of 1943, Walter Pitts solves the inverse network problem, "which is, given a preassigned pattern of activity over time, to construct when possible a neuron-network having this pattern" [34, p. 23] by allowing creating problem-oriented neural networks. Tara H. Abraham indicates that adopting Householder's model of neural excitation, Walter Pitts develops a simpler procedure for the mathematical analysis of excitatory and inhibitory activity in a simple neuron circuit, and aimed to develop a model applicable to the most general neural network possible [2].

"Psychometrika", the official journal of the Psychometric Society (both founded in 1935 by Louis Leon Thurstone, Edward Lee Thorndike and Joy Paul Guilford), is devoted to the development of psychology as a quantitative rational science. It has become another mouthpiece of Nicolas Rashevsky and his students, whose articles examine statistical methods, discuss mathematical techniques, and advance theory for evaluating behavioral data in psychology, education, and the social and behavioral sciences generally. Walter Pitts's article "A general theory of learning and conditioning" has been published in this journal. "The field of conditioning and learning has attained a development on the purely experimental side which renders it an excellent point for the entry of quantitative theory into psychology... The work of this kind done so far has been chiefly from two standpoints: the first ... attempts with some success to explain the phenomena directly upon a neurological basis, while the second ... prefers to elaborate first a macroscopic account of behavior per se, while leaving the neurological foundations until a later stage. These two approaches are of course rather complementary than competitive: the development of theoretical neurology provides very many suggestions for macroscopic work, and the latter simplifies the neurological problem by requiring mechanisms to account for only a few general propositions instead of a multitude of facts in no obvious relation. ... We shall consider the results of our discussion applicable to all aspects of learning and conditioning in which the effect of symbolic or verbal factors is not of great significance; and within this field we shall deal with all eases of learning and conditioning in which independent or related stimuli, with given original tendencies to produce specified types of response, are distributed over time in specified intensities in an arbitrary way, continuous or otherwise; and in which affective stimulation, if this form part of the experimental routine, is distributed in any given manner. In partial confirmation of our hypotheses, we shall point out how most of the principal experimental generalizations can be inferred from the theory, at least as regards comparative order of magnitude; while a rigid quantitative test would require data in a detail not ordinarily given in experimental results. The theory does not seem too difficult to verify in most of its aspects, however, by a fairly extended and precise set of experiments, whose performance would also provide direct information upon a number of matters of considerable import upon which little data are available, and which, even if disconfirming our present system in some of its aspects, would assuredly make suggestions leading to a better one of comparable range and generality." [31, pp. 1-3]

Part I [31] deals only with the case where the stimuli and responses are wholly independent, so that transfer and generalization do not occur, and proposes a law of variation for the reaction-tendency, which takes into account all of classical conditioning and the various sorts of inhibition affecting it. Part II [32] extends a mathematical theory of non-symbolic learning and conditioning, still under the hypothesis of complete independence, to cases where reward and punishment are involved as motivating factors. The preceding results are generalized to the case where stimuli and responses are related psychophysically, thus constituting a theory of transfer, generalization, and discrimination.

Another article of 1943, "A logical calculus of the ideas immanent in nervous activity" [26], published again in "Bulletin of Mathematical Biophysics", has resulted from cooperation of Warren Sturgis McCulloch and Walter Pitts and is considered one of the most famous papers on artificial neural networks. They stated five physical assumptions for nets without circles [26, p. 118]:

1. The activity of the neuron is an "all-or-none" process [any nerve has a finite threshold and the intensity of excitation must exceed this for production of excitation – once produced, the excitation proceeds independently of the intensity of the stimulus].
2. A certain fixed number of synapses must be excited within the period of latent addition [time during which the neuron is able to detect the values present on its inputs, the synapses – typically less than 0.25 msec] in order to excite a neuron at any time, and this number is independent of previous activity and position on the neuron.

3. The only significant delay within the nervous system is synaptic delay [time delay between sensing inputs and acting on them by transmitting an outgoing pulse, – typically less than 0.5 msec].
4. The activity of any inhibitory synapse absolutely prevents excitation of the neuron at that time.
5. The structure of the net does not change with time.

The neuron described by these five assumptions is known as the McCulloch-Pitts neuron [7, p. 17]. In the same way as propositions in propositional logic can be "true" or "false," neurons can be "on" or "off" – they either fire or they do not: this formal equivalence allowed them to argue that the relations among propositions can correspond to the relations among neurons, and that neuronal activity can be represented as a proposition [2, p. 19]. "In this way all nets may be regarded as built out of the fundamental elements of Figures a, b, c, d, precisely as the temporal propositional expressions are generated out of the operations of precession, disjunction, conjunction, and conjoined negation. In particular, corresponding to any description of state, or distribution of the values *true* and *false* for the actions of all the neurons of a net save that which makes them all false, a single neuron is constructible whose firing is a necessary and sufficient condition for the validity of that description. Moreover, there is always an indefinite number of topologically different nets realizing any temporal propositional expression" [26, p. 121].

Fig. 5 reveals fundamental elements of McCulloch-Pitts neural networks. The triangle depicts the neuron body, figures in the triangles are neuron numbers, lines are axons, dots adjacent to neurons are excitatory synaptic connections, open circles adjacent to the neuron are inhibitory synaptic connections. In author's expressions for the figures, the dots on either side of the "≡" symbol act as separators, ≡ act as biconditional logical connectives (logical equivalence), single dots act as conjunction (logical AND), ∨ act as disjunction (logical OR), and ~ act as negation (logical NOT).

In Fig. 5(a), neuron 2 will fire if and only if neuron 1 fires. Logically, this corresponds to the expression $N_2(t) \Leftrightarrow N_1(t-1)$, which can be read as "neuron 2 will fire at time ($t$) if and only if neuron 1 fires at time ($t-1$)". Fig. 5(b) shows a network that is isomorphic with the Boolean function "OR" in propositional logic. Its expression, $N_3(t) \Leftrightarrow N_1(t-1) \vee N_2(t-1)$ means that neuron 3 will fire at time ($t$) if and only if neuron 1 fires or neuron 2 fires at time ($t-1$). Fig. 5(c) demonstrates the Boolean "AND" function. The expression $N_3(t) \Leftrightarrow N_1(t-1) \wedge N_2(t-1)$ means that neuron 3 will fire at time ($t$) if and only if neuron 1 fires at time ($t-1$) and neuron 2 fires at time ($t-1$). Warren Sturgis McCulloch and Walter Pitts don't provided an example of the 'clean' Boolean "NOT" function – instead it they use a conjoined negation with the instance of an inhibitory neuron. The logical expression $N_3(t) \Leftrightarrow N_1(t-1) \wedge \neg N_2(t-1)$ corresponding to Fig. 5(d), means that neuron 3 will fire at time ($t$) only if neuron 1 fires at time ($t-1$) and neuron 2 does not fire at time ($t-1$).

In [26], there is a set of theorems that "does in fact provide a very convenient and workable procedure for constructing nervous nets to order, for those cases where there is no reference to events indefinitely far in the past in the specification of the conditions" [26, pp. 121–122]. Warren Sturgis McCulloch and Walter Pitts appear to be the first authors since William James to describe a massively parallel neural model. The theories they developed were important for a number of reasons, including the fact that any finite logical expression can be realized by networks of their neurons.

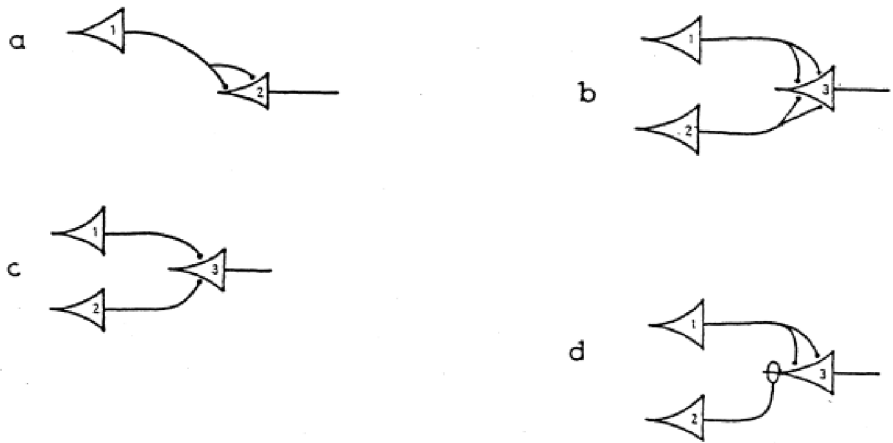

EXPRESSION FOR THE FIGURES

a  $N_2(t) . \equiv . N_1(t-1)$
b  $N_3(t) . \equiv . N_1(t-1) \vee N_2(t-1)$
c  $N_3(t) . \equiv . N_1(t-1) . N_2(t-1)$
d  $N_3(t) . \equiv . N_1(t-1) . \sim N_2(t-1)$

**Fig. 5.** Fundamental elements of McCulloch-Pitts neural networks

Combining simple "logical" neurons in chains and cycles, the authors show that the brain is able to perform any logical operation and arbitrary logical calculations. The paper is essential for developing computing machines as it allows creating a universal computer operating with logical expressions (in the hands of John von Neumann, the McCulloch-Pitts model becomes the basis for the logical design of digital computers [6, p. 180]): "It is easily shown: first, that every net, if furnished with a tape, scanners connected to afferents, and suitable efferents to perform the necessary motor-operations, can compute only such numbers as can a Turing machine; second, that each of the latter numbers can be computed by such a net; and that nets with circles can be computed by such a net; and that nets with circles can compute, without scanners and a tape, some of the numbers the machine can, but no others, and not all of them. This is of interest as affording a psychological justification of the Turing definition of computability and its equivalents, Church's λ-definability and Kleene's primitive recursiveness: If any number can be computed by an organism, it is computable by these definitions, and conversely." [26, pp. 121–122]

In the same issue of "Bulletin of Mathematical Biophysics", in which [26] was published, Herbert Daniel Landahl (the first doctoral student in Nicolas Rashevsky's mathematical biology program at the University of Chicago, who became the second President of the Society for Mathematical Biology in 1981), Warren Sturgis McCulloch and Walter Pitts published a short (about 3 pages), yet essential addition [23], suggesting a method for converting logical relations among the actions of neurons in a net into statistical relations among the frequencies of their impulses. In the presented theorem, they detailed transition from Boolean calculations (in "true" and "false") to probabilistic ones (numbers within [0; 1]): the conjunction sign $\vee$ is replaced by $+$, the disjunction sign (single dot) is replaced by $\times$, negation $\sim$ is replaced by «1 –», etc. The correspondence expressed by this theorem connects the logical calculus of the [26] with previous treatments of the activity of nervous nets in mathematical biophysics and with quantitatively measurable psychological phenomena.

The monograph by Alston Scott Householder and Herbert Daniel Landahl "Mathematical Biophysics of the Central Nervous System" has become a kind of conclusion of the discussed period [12]. In Paul Cull's opinion, there is no unambiguous answer to the question which model is better, the Rashevsky continuous model or the McCulloch-Pitts discrete model: "For some purposes, one model is better, but for other purposes, the other model is better. Rashevsky and Landahl were quick to notice, that in physics, one often averaged over a large set of discrete events to obtain a continuous model which represented the large scale behavior of a system, and so they posited that the continuous neuron model might be suitable for modeling whole masses of neurons even if each individual neuron obeyed a discrete model. In the hands of Householder and Landahl, this observation led to the idea of modeling psychological phenomena by neural nets with a small number of continuous model neurons. In particular, they found that the cross-couple connection [Fig. 6] was extremely useful. For such problems as reaction time, enhancement effects, flicker phenomena, apparent motion, discrimination and recognition, they were able to fit these models to experimental data and to use their models to predict phenomena that could be measured and verified" [6, p. 180].

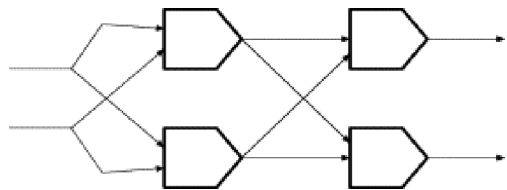

**Fig. 6.** This cross-couple connection of four neurons is capable of modeling a large number of phenomena (according to [6])

In 1945, Nicolas Rashevsky wrote about [26] and [23]: "authors show that by applying logical calculus, it is possible to construct any complicated network having given properties. One could attempt to construct by the method of McCulloch and Pitts a network that would represent all modes of logical reasoning, and then apply the usual methods of mathematical biophysics to derive some quantitative relations between different manifestations of the processes of logical thinking" [43, p. 146]. "It seems somewhat awkward to have to construct by means of Boolean algebra first a "microscopic circuit" and then obtain a simpler one by a transition to the "macroscopic" picture. We should expect that a generalization of the application of Boolean algebra should be possible so as to permit its use for the construction of networks in which time relations are of a continuous rather than of a quantized, nature" [45, p. 211].

Nicolas Rashevsky intensively develops the apparatus created by McCulloch and Pitts in his further papers. In [46] a theory of such neural circuits is developed which provide for formal logical thinking. As a by-product of this study, a neural mechanism is indicated which provides for the conception of ordinal numbers. A quantitative theory of the probability of erroneous reasoning and of the speed of reasoning in its relations to other psychological phenomena is suggested. Predicate apparatus application enables Nicolas Rashevsky synthesizing huge neural networks from single-type fundamental elements of McCulloch-Pitts (Fig. 7).

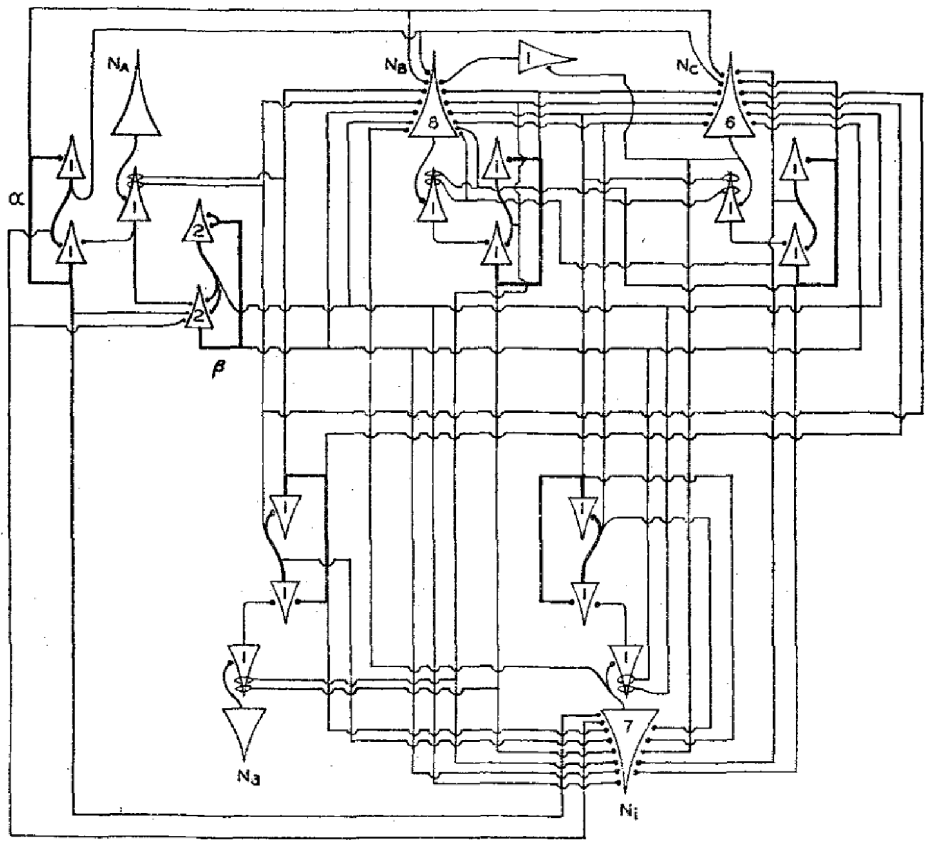

**Fig. 7.** Nicolas Rashevsky's complex neural network example (from [46, p. 32])

In their paper of 1946, Herbert Daniel Landahl and Richard Runge [24] make the next move towards spreadsheet interpretation of neural networks: the activity of a neural net is represented in terms of a matrix vector equation with a normalizing operator in which the matrix represents only the complete structure of the net, and the normalized vector-matrix product represents the activity of all the non-afferent neurons.

Let a net of $n$ neurons be divided into afferents, efferents, and internal neurons. We shall only mean by an afferent a neuron not acted upon by any neuron in the net under consideration. Similarly, an efferent is a neuron which does not act on any other neuron in the net. We may, however, refer to the afferents as receptors and the efferents as effectors. Let the $\rho$ receptor neurons be $N_1, N_2, ..., N_r, ..., N_\rho$, the $\iota$ internals be $N_{\rho+1}, ..., N_i, ..., N_{\rho+\iota}$ and the $\varepsilon$ effector neuron be $N_{n-\varepsilon+1}, ..., N_e, ..., N_n$. Define a structure matrix $F$ as a square matrix having the row and column indices 1, 2, ..., $n$ corresponding to the $n$ neurons such that

$$F = |f_{jk}| = \begin{array}{c|c|c|c|} & 1...r...\rho & \rho+1...i...\iota & (n-\varepsilon+1)...e...n \\ \hline 1 & 0 & F_R & F_X \\ \hline & 0 & F_I & F_E \\ \hline n & 0 & 0 & 0 \\ \hline \end{array} \quad (2)$$

Each element in row $j$ determines which neurons are acted upon by neuron $N_j$ and in what manner. Similarly each element in column $k$ determines which neurons act upon $N_k$ and in what manner. The matrices $F_R$, $F_X$, $F_I$, and $F_E$, appearing in equation (2), determine respectively the relationships receptor-internal, receptor-effector, internal-internal, and internal-effector. Define the matrix $R$ as the $n \times n$ matrix, obtained from $F$ by substituting zeros for all elements except those of $F_R$. Define in a similar way the matrices $X$, $I$, and $E$. We shall assume that no neuron acts upon itself so that $f_{kk} = 0$ for all $k$, and $F$ and $F_I$ have all diagonal elements equal to zero.

If any row $\alpha$ in $F$ contains only positive or zero elements, then $N_\alpha$ is a purely excitatory neuron. If any row $\beta$ in $F$ contains only negative or zero elements, then $N_\beta$ is a purely inhibitory neuron. If both positive and negative elements occur in a given row, the corresponding neuron may be referred to as a mixed neuron.

Define the $(1 \times n)$ row matrix or vector $\boldsymbol{a}$, by

$$\boldsymbol{a}(t) = (a_1, ..., a_r, ..., a_\rho, a_{\rho+1}, ..., a_i, ..., a_\iota, a_{n-\varepsilon+1}, ..., a_e, ..., a_n), \quad (3)$$

where any element $a_j$ is 1 or 0 depending on whether $N_j$ does or does not act at the time $t$. The vector $\boldsymbol{a}(t)$ may be referred to as the activity vector at the time $t$. This vector may be written as the sum of three $(1 \times n)$ vector components, $\boldsymbol{r}, \boldsymbol{i}, \boldsymbol{e}$, the receptor, internal, and effector components having the respective set of elements $a_r$, $a_i$, and $a_e$ only, and zeros elsewhere. The scalar quantity $v_k(t)$ given by the sum

$$v_k(t) = \sum_{j \epsilon \beta} f_{jk} = \sum_{j=1}^{N} a_j(t-1) f_{jk}, \qquad (4)$$

taken over a class $\beta$ of neurons which is defined as the class of all neurons synapsing on $N_k$ which are active at $t-1$, gives a measure of the net excitation affecting the neuron $N_k$. A vector $v(t)$, whose components for $k > \rho$ are the values of $v_k(t)$ and for $k \leq \rho$ – the values of $a_k(t)$, can be expressed as

$$v(t)=r(t)+a(t-1)F. \qquad (5)$$

In order to normalize $v(t)$, let $\mathcal{G}$ be a post-operator on a row vector, such that, if $[v\mathcal{G}]_k$ is the $k$th component of the vector $[v\mathcal{G}](t)$,

$$[v\mathcal{G}]_k = \begin{cases} 1 \text{ if } v_k \geq 1, \\ 0 \text{ if } v_k < 1. \end{cases} \qquad (6)$$

Since $\mathcal{G}$ operates only on a vector appearing just to the left of it, and not on a matrix, the expression $(vF)\mathcal{G}$ may be written $vF\mathcal{G}$ (the parenthesis could not be eliminated if a pre-operator were used). Equation (5) may be written

$$a(t)=v(t)\mathcal{G}=[r(t)+a(t-1)F]\mathcal{G}. \qquad (6)$$

According to the given theorem [61, p. 78], the activity of any net represented by a structure matrix $F$ is determined from the afferent stimulation and its activity at the beginning of the prior interval of time according to the equation

$$a(t)=r(t)+a(t-1)F\mathcal{G}. \qquad (7)$$

Equation (8) leads to the recursion formula

$$a(t)=r(t)+\{r(t-1)+[r(t-2)+... + [r(2)+[r(1)+a(0)]F\mathcal{G}]F\mathcal{G}]... F\mathcal{G}]F\mathcal{G}\}F\mathcal{G}. \qquad (8)$$

The vector quantity between the first and last brackets in expression (9) is simply the vector $a(t-1)$.

Thus if the structure of a net is known together with a sequence of $r$'s, $r(0)$, $r(1)$, ..., $r(\tau-1)$, and the initial activity of the internal neurons $i(0)$, it is possible from equation (9) to determine the activities of the net for any time $t$ from 0 to $\tau$, $a(1)$, $a(2)$, ..., $a(\tau)$.

Because of the character of the structure matrix F, equation (8) may be written as a pair of equations

$$i(t)=[r(t-1)R+i(t-1)I]\mathcal{G}, \qquad (9)$$

$$e(t)=[r(t-1)X+i(t-1)E]\mathcal{G}, \qquad (10)$$

from which one may determine successively $i(t)$ and $e(t)$ for every $t$. For both $i(t)$ and $e(t)$ formulas similar to expression (9) can be written.

From equation (10), it is evident that the sequence of $i$'s can be determined from a knowledge of the afferent-internal structure and internal-internal structures, together with the sequence of afferent activities and an initial internal activity pattern. On the other hand, to determine the sequence of *e*'s, that is, the pattern of the efferent activity, one must also know the rest of the structure of the net.

Every row and column of *F*, excluding the first $\rho$ columns and last $\varepsilon$ rows, contains at least one non-zero entry; otherwise, it represents an afferent or efferent. If only one non-zero entry occurs in any column, it may be replaced by unity, for if it is less than one, the neuron of the corresponding column can never act, and thus this neuron should be deleted. Furthermore, there is no restriction to set an element equal to one, if this element is greater than one.

Authors give an example of the application of the presented in [24] method – the matrix *F* for the circuit illustrated in Fig. 8.

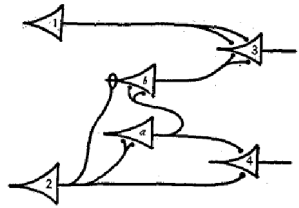

**Fig. 8.** Figure 1e of the paper "A logical calculus of the ideas immanent in nervous activity" [26]

$$F = \begin{array}{c|c|cc|cc} & 1\ 2 & a\ \ b & 3\ \ 4 \\ \hline 1 & & 0\ \ 0 & 1\ \ 0 \\ 2 & 0 & 1\ -1 & 0\ \ \tfrac{1}{2} \\ \hline a & & 0\ \ 1 & 0\ \ \tfrac{1}{2} \\ b & 0 & 0\ \ 0 & 1\ \ 0 \\ \hline 3 & & & \\ 4 & 0 & 0 & 0 \end{array}$$

If the net is initially at rest, then, according to [26], the expression for neuron 3 is $N_3(t) \Leftrightarrow N_1(t-1) \vee N_2(t-3) \wedge \neg N_2(t-2)$ and the condition for the activity of neuron 3 at $t=0$, is either that neuron 2 acts at $t=-3$ but not at $t=-2$; or that neuron 1 acts at $t=-1$. That is, the sequence of *r*'s, writing only the $a_r$, components, $r(-3)=(0, 1)$, $r(-2)=(0, 0)$, $r(-1)=(0, 0)$ as well as the sequence $r(-1)=(1, 0)$ is adequate to produce activity in neuron 3. Assume that $r(t)=0$ for all $t \geq 0$. If $r(-3)=(0, 1)$ then from equation (8) $a(-2)=(0, 0; 1, -1; 0, \tfrac{1}{2})\mathcal{G}=(0, 0; 1, 0; 0, 0)$, $a(-1)=(0, 0; 0, 1; 0, 0)$, and $a(0)=(0, 0; 0, 0; 1, 0)$, so that $e(0)$, writing only the $a_e$ components is $e(0)=(1, 0)$, that is, neuron 3 acts but 4 does not. Similarly if $r(-1)=(1, 0)$ then from equation (8) $e(0)=(1, 0)$. For both sequences $a(1)=0$.

The expression for neuron 4 is $N_4(t) \Leftrightarrow N_2(t-2) \wedge N_2(t-1)$, the condition for neuron 4 to act as $t=0$ is that neuron 2 acts at $t=-2$ and at $t=-1$. If $r(-2)=(0, 1)$ and $r(-1)=(0, 1)$, then from equation (8) $a(-1)=(0, 1; 1, 0; 0, 0)$ and $a(0)=(0, 0; 1, 0; 0, 1)$, so that neuron 4 acts but 3 does not. For this sequence $a(1)=(0, 0; 0, 1; 0, 0)$ and $a(2)=(0, 0; 0, 0; 1, 0)$ so that neuron 3 always acts as a unit of time after discontianuation of continuous stimulation of neuron 2.

In a paper [24] a method was given by which the efferent activity of an idealized neural net could be calculated from a given afferent pattern. Those results are extended in the next-year paper [25]: (1) conditions are given under which nets may be considered equivalent, (2) rules are given for the reduction or extension of a net to an equivalent net, (3) a procedure is given for constructing a net which has the property of converting each of a given set of afferent activity patterns into its corresponding prescribed efferent activity pattern.

Telson Wei develops another approach to matrix representation of a neural network [63]. The structure of a complete or incomplete neural net is represented here by several matrices: the intensity matrix $E$, the connection matrix $D$, the structural matrix $T$, the diagonal inverse threshold-matrix $H$, and activity vector $\boldsymbol{a}$ from [24; 25]. The activity equation of the net follows in a general form. A chain or cycle is defined as a neural structure whose connection matrix is unitary. Telson Wei computes the number of simple chains by a recurrent formula.

In their paper of 1948 [51], Alfonso Shimbel and Anatol Rapoport (pioneered in the modeling of parasitism and symbiosis, researching cybernetic theory) develop a probabilistic approach to the theory of neural nets: neural nets are characterized by certain parameters which give the probability distributions of different kinds of synaptic connections throughout the net. In their further papers, they consider steady states in random nets [37; 42] and contribution to the probabilistic theory of neural nets: randomization of refractory periods and of stimulus intervals [38], facilitation and threshold phenomena [39], specific inhibition [40] and various models for inhibition [41].

The last joint article by Walter Pitts and Warren Sturgis McCulloch, "How we know universals the perception of auditory and visual forms", in "Bulletin of Mathematical Biophysics" came out in 1947. "Numerous nets, embodied in special nervous structures, serve to classify information according to useful common characters. In vision they detect the equivalence of apparitions related by similarity and congruence, like those of a single physical thing seen from various places. In audition, they recognize timbre and chord, regardless of pitch. The equivalent apparitions in all cases share a common figure and define a group of transformations that take the equivalents into one another but preserve the figure invariant. So, for example, the group of translations removes a square appearing at one place to other places; but the figure of a square it leaves invariant. ... We seek general methods for designing nervous nets which recognize figures in such a way as to produce the same output for every input belonging to the figure. We endeavor particularly to find those which fit the histology and physiology of the actual structure." [30, pp. 127–128]

Two neural mechanisms are described which exhibit recognition of forms. Both are independent of small perturbations at synapses of excitation, threshold, and synchrony, and are referred to particular appropriate regions of the nervous system, thus suggesting experimental verification. The first mechanism averages an apparition over a group, and in the treatment of this mechanism it is suggested that scansion plays a significant part. The second mechanism reduces an apparition to a standard selected from among its many legitimate presentations. The former mechanism is exemplified by the recognition of chords regardless of pitch and shapes regardless of size. Both are extensions to contemporaneous functions of the knowing of universals heretofore treated by the authors only with respect to sequence in time.

"We have focused our attention on particular hypothetical mechanisms in order to reach explicit notions about them which guide both histological studies and experiment. If mistaken, they still present the possible kinds of hypothetical mechanisms and the general character of circuits which recognize universals, and give practical methods for their design. These procedures are a systematic development of the conception of reverberating neuronal chains, which themselves, in preserving the sequence of events while forgetting their time of happening, are abstracted universals of a kind. Our circuits extend the abstraction to a wide realm of properties. By systematic use of the principle of the exchangeability of time and space, we have enlarged the realm enormously. The adaptability of our methods to unusual forms of input is matched by the equally unusual form of their invariant output, which will rarely resemble the thing it means any closer than a man's name does his face." [30, p. 146]

Thus, the models and methods developed by Walter Pitts and Warren Sturgis McCulloch have created a foundation for designing a new type of computers – neurocomputers based on human brain principles and able to solve tasks of recognizing distorted (noisy) images.

## 5      Conclusions

1. Extensive application of artificial intelligence in everyday life calls for students' early acquaintance with its models and methods including neural network-based while teaching informatics at secondary schools. It conditions the need for developing training methods of computer simulation of neural networks in the general-purpose simulation environment, i.e. spreadsheets.
2. Basic solutions of the problem of computer simulation training of neural networks in the spreadsheet environment include: 1) joint application of spreadsheets and network simulation tools; 2) application of third-party add-ins to spreadsheet processors; 3) macros development using embedded languages of spreadsheet processors; 4) application of standard spreadsheet add-ins for non-linear optimization; 5) creation of neural networks in the spreadsheet environment without add-ins and macros.
3. Neural network simulation competences should be formed through mastering models based on the historical and genetic approach. The review of papers on computational neuroscience of its early period allows determining three groups of models, which are helpful for developing corresponding methods: the continuous two-factor model of Rashevsky, the discrete model of McCulloch and Pitts, and the discrete-continuous models of Householder and Landahl.
4. Further research implies considering mathematical models of the Age of Camelot and developing their spreadsheet interpretations of various complexity.